\documentclass[twoside]{article}

\usepackage[round]{natbib}
\usepackage[nottoc,numbib]{tocbibind}
\usepackage{makecell}
\usepackage{graphicx}

\usepackage[accepted]{arxiv}

\begin{document}

\twocolumn[

\aistatstitle{Latent Vector Recovery of Audio GANs}

\aistatsauthor{ Andrew Keyes* \And Nicky Bayat* \And  Vahid Reza Khazaie \And Yalda Mohsenzadeh }

\aistatsaddress{ Western University \\ London, ON, Canada \\ akeyes6@uwo.ca  \And Western University \\ London, ON, Canada \\ nbayat5@uwo.ca \And Western University \\ London, ON, Canada \\ vkhazaie@uwo.ca \And Western University \\ London, ON, Canada \\ ymohsenz@uwo.ca } ]

\begin{abstract}
Advanced Generative Adversarial Networks (GANs) are remarkable in generating intelligible audio from a random latent vector. In this paper, we examine the task of recovering the latent vector of both synthesized and real audio. Previous works recovered latent vectors of given audio through an auto-encoder inspired technique that trains an encoder network either in parallel with the GAN or after the generator is trained. With our approach, we train a deep residual neural network architecture to project audio synthesized by WaveGAN into the corresponding latent space with near identical reconstruction performance. To accommodate for the lack of an original latent vector for real audio, we optimize the residual network on the perceptual loss between the real audio samples and the reconstructed audio of the predicted latent vectors. In the case of synthesized audio, the Mean Squared Error (MSE) between the ground truth and recovered latent vector is minimized as well. We further investigated the audio reconstruction performance when several gradient optimization steps are applied to the predicted latent vector. Through our deep neural network based method of training on real and synthesized audio, we are able to predict a latent vector that corresponds to a reasonable reconstruction of real audio. Even though we evaluated our method on WaveGAN, our proposed method is universal and can be applied to any other GANs.
\end{abstract}

\section{INTRODUCTION}
Researchers have recently shown an increased interest in mapping generated samples by generative adversarial networks (GANs) into the original latent vectors. GANs consist of two major components: the generator and the discriminator. The generator receives a random latent vector and generates realistic samples through a min-max optimization game with the discriminator. The discriminator aims to determine whether a sample is real or fake. The competition between the two networks helps to enhance their respective performances. After training, for a given random vector z, GANs generate realistic samples. The problem of GAN inversion refers to the task of recovering a latent vector for a given sample (image, audio, video, etc.) that when given to the generator, it can precisely regenerate the target.

Inverse mapping of GANs has a pivotal role in a wide range of scientific and industrial applications. This inversion is of interest because it can be used as a measure for comparing GANs' performances \citep{creswell2018inverting}. If a GAN is capable of generating samples that can be mapped to more precise recovered latent vectors, it has learned richer information from the training distribution and thus has a better performance compared to the other GANs. Furthermore, since modifying latent vectors results in meaningful changes in the image/audio domain \citep{radford2015unsupervised}, the inverse mapping of GANs can be used in classification or retrieval problems. By transforming the latent vector of an image/audio, one can add styles or features to the generated sample. For example, it is possible to generate more memorable images with GANs by linearly transforming the latent representation \citep{goetschalckx2019ganalyze}. One possible example of style transfer in audio is changing the speaker identifier of audio while maintaining the content of the speech \citep{van2017neural}.

Projecting generated samples back to the latent-space is a classic problem in computer vision, where the goal is to recover the latent vector of given images. However, investigating the inverse mapping of GANs is still a continuing concern within the audio domain. While methods introduced in the vision domain can be applied to audio as well, they need to be improved and fine-tuned to this domain in order to produce ideal results.

Four major solutions have been proposed for projecting images back to the latent vector peers. There is a large and growing body of the literature that recovers the latent vector through an optimization-based approach \citep{creswell2018inverting}. This technique updates an initial random z vector using gradient descent until the loss between the sample generated by this vector and the target is less than a desired threshold. Auto-encoder inspired architectures are the second most common solution to this problem \citep{van2017neural}\citep{donahue2016adversarial}\citep{dumoulin2016adversarially}. A third encoder network is trained to map generated samples to latent vectors. Next, the generator (decoder) uses these vectors to regenerate the target sample.

Deep neural networks are a recent tool for learning the inverse mapping of GANs with high fidelity and speed. They can be trained on a dataset of generated samples and matching latent vectors \citep{bayat2020inverse}. Hybrid methods are another approach that benefits from the advantages of both mentioned techniques \citep{zhu2016generative}. They first predict an initial latent vector via a deep network and later update via a few further steps of gradient descent.

It has previously been observed that it is possible to map synthesized images to latent vectors that regenerate indistinguishable results from the target. The main challenge faced by many researchers is the lack of ground truth latent vectors for real samples. In this paper, we propose a residual deep neural network based method that predicts latent vectors of audio samples using a ResNet-18 architecture. Our inverse mapping model is trained based on a combination of pixel and perceptual loss. Later, we demonstrate our results on latent vector recovery of both generated and real audio samples and compare them with sole gradient descent optimization and hybrid methods that apply a few steps of gradient descent after the initial prediction by the deep neural network methods.

To the best of our knowledge, the experimental work presented here provides one of the first investigations into how to project audio samples to latent representations that can regenerate them. In this paper our contributions are the following:
\begin{itemize}
    \item We propose a fast deep network technique to recover latent vectors of both fake and real audio.
    \item We propose a novel perceptual loss within the audio domain.
    \item We also perform latent vector recovery using gradient descent methods in the audio domain.
    \item We implement a hybrid method that applies a few steps of gradient descent after predicting the representation by the deep network and compare the results with the proposed ResNet-18 framework and sole gradient descent approach.
\end{itemize}

\section{PREVIOUS WORKS}
Recent advances in the inverse mapping of adversarial generators, in particular inverse mapping of image generators, have led to the question of whether it is possible to map audio to a latent vector that can regenerate it. Despite its many applications, there has been no detailed investigation on recovering the latent vectors of audio. This paper attempts to show that common approaches used for projecting images to latent-space can be used on audio, but they need to be fine-tuned and adjusted for this task.
There are four major approaches proposed to invert the generators of GANs.

\begin{figure*}[t]
\centerline{\includegraphics[width=1\linewidth]{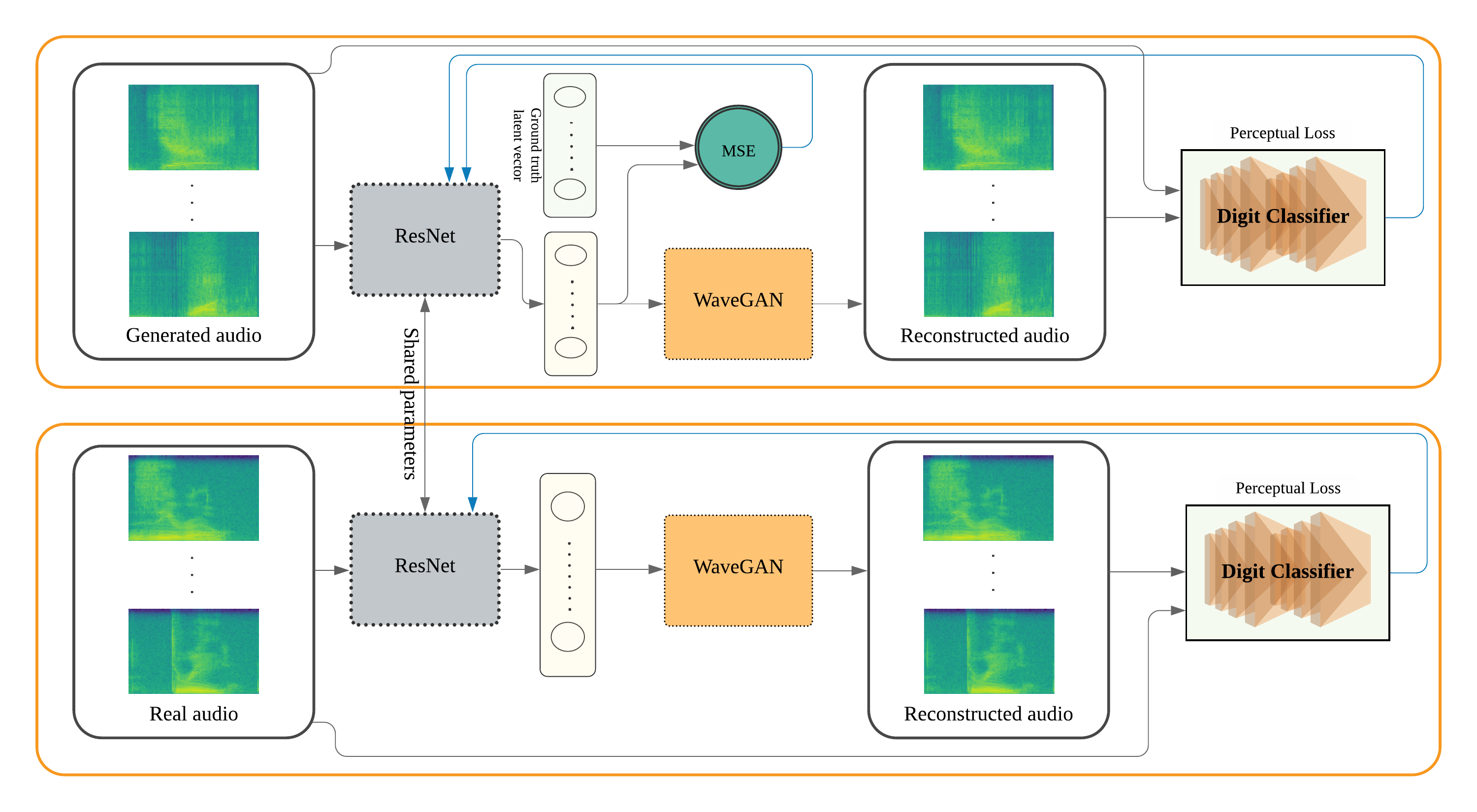}}
\caption{The Two Branch Architecture of Our Deep Network Based Inversion Technique for Real Audio. \label{fig1}}
\end{figure*}

There is a large volume of published studies on gradient-based inversion methods \citep{creswell2018inverting}. In order to find the optimal latent vector, they start from a random vector $z^*$ sampled from the training distribution. At every iteration, they update this representation to minimize the loss between the generated sample $G(z^*)$ and the target until the loss is below a certain threshold. While this method can recover accurate latent vectors, it often takes too long to find the solution. \citep{lipton2017precise} proposed the stochastic clipping technique to bound the latent vectors to the original distribution while being updated by gradient descent. Any value larger than the maximum number allowed or smaller than the minimum, is replaced by a random number within the acceptable range.

A considerable amount of literature has been published on auto-encoder inspired architectures \citep{van2017neural}\citep{donahue2016adversarial}\citep{dumoulin2016adversarially}. They learn the inverse mapping by training an encoder either at the same time as the GAN or after the GAN is trained. The encoder maps generated samples back to the latent domain. The approach of training an encoder alongside the GAN cannot be used for pre-trained GANs. In addition, adding a third network adds extra parameters to the training which might cause over-fitting.

More recent attention has focused on using deep neural networks to predict latent vectors. \citep{bayat2020inverse} proposed a deep residual network (ResNet-18) to learn the inverse mapping based on both reconstruction and perceptual loss. Pixel loss between predicted and ground truth latent vectors is used as the reconstruction loss. The perceptual loss between the reconstructed and target images are computed based on Mean Squared Error (MSE) between features extracted from concatenation layers of a pre-trained FaceNet \citep{schroff2015facenet}.

A number of studies have begun to examine a hybrid method of inverting GANs \citep{zhu2016generative}. Using this approach, researchers have been able to benefit from the advantages of both optimization and deep neural network based methods. This approach first predicts the latent vector using a pre-trained deep neural network, then fine-tune the prediction using a few steps of gradient descent. Even though this method works well on images, we demonstrate that sole deep network predictions perform better than hybrid optimizations.

In this paper, we build on \citep{bayat2020inverse}, and employ the framework to recover latent vectors for synthetic and real audio. We compare our method with both optimization-based and hybrid inversion methods and show our superiority.

\section{METHOD}
A number of techniques have been developed for recovering latent vectors of images. Including optimization-based methods, auto-encoder inspired architectures, and deep neural network models. A major advantage of using deep residual neural networks is that they are incredibly fast when compared to gradient-based alternatives. The second advantage of using the ResNet architecture is that we can train it on a dataset of generated audio and their corresponding ground truth latent vector to learn the inverse mapping.  In this work, the ResNet-18 architecture previously proposed for image latent vector recovery \citep{bayat2020inverse} is assayed for audio domain using generated audio and matching latent representation as well as a novel perceptual loss for audio.

\subsection{WaveGAN}
WaveGAN was the first attempt at applying GANs to unsupervised synthesis of raw-waveform audio \citep{wavegan}. The model is capable of producing one second clips at a frequency of 16khz learned from a variety of audio datasets. WaveGAN makes use of transpose convolutions to upsample from a latent space to the final raw-waveform audio. To avoid checkerboard artifacts in audio, a phase shuffle method is applied to perturb the phase of the discriminator's activations by -n to n. We chose to use the official WaveGAN model pre-trained on the SC09 (subset of Speech Commands) dataset for two reasons \citep{warden2018speech}. First from a qualitative perspective, it was much easier to distinguish whether the style and content of a reproduced spoken digit remained intact rather than drums or bird calls. Second, the perceptual loss could be taken from feature maps of a pre-trained classifier on the SC09 dataset. Also, predicting the latent vector of real speech can be useful for applying style or content transformations through a transformer in the latent space. 

\subsection{Inverse Mapping Model}
Our goal for the inverse mapping model was to take as input the spectrogram of generated and real audio and output the predicted latent vector for a pre-trained WaveGAN. The inverse mapping model architecture is the same as ResNet-18 with the last activation function removed and the number of classes set to the size of the WaveGAN latent space. The ResNet-18 architecture was selected as it has shown to be useful in the recovery of latent vectors in face GANs \citep{bayat2020inverse}. At each epoch, the inverse mapping model was trained equally on audio synthesized by WaveGAN at run time from a random latent vector and real audio from the SC09 dataset. The architecture of our proposed method is presented in Figure \ref{fig1}.

In the case of synthesized audio, two losses were used to train the network. The first was the MSE between the original latent vector used as input to the WaveGAN and the predicted latent vector by the inverse mapping model. Second, the perceptual loss of the audio synthesized by the original latent vector and the reconstructed audio from the predicted latent vector was calculated through a classifier model trained on predicting the spoken digit. For real audio, only the perceptual loss was used as we do not have a true latent vector. The MSE loss in combination with the perceptual loss helped to learn the inverse mapping from real and synthesized audio to the latent space while maintaining content and style. 

\subsubsection{Objective Function}
To train the inverse mapping model, we employed two types of objective functions:
\begin{itemize}
    \item \textbf{MSE between latent vectors:} At run time, audio was synthesized by WaveGAN from a random latent vector with a uniform distribution between -1 and 1. With the synthesized audio as input to the inverse mapping model, the latent vector was predicted in the last layer of the network. The ground truth latent vector was then compared to the predicted latent vector through mean squared error. This loss is necessary to learn the inverse mapping between the audio and latent vectors. 
    
    \item \textbf{Perceptual loss:} For real and synthesized audio, the perceptual loss was back-propagated through the inverse mapping model. The perceptual loss was quantified as the difference in activations in a ResNet-18 spoken digit classifier's output at each residual block. The classifier was pre-trained on the SC09 dataset.  This loss improved the reconstruction of real audio and helped keep style and content intact. 
\end{itemize}

\begin{table*}[h]
\caption{Synthesized Audio Reconstructions} \label{fake_audio_table}
\begin{center}
\begin{tabular}{|c|c|c|c|}
\hline
\textbf{}  & \makecell{Inception Score \\ (mean $\pm$\ std)} &  \makecell{MSE \\ (raw audio)} & \makecell{SSIM \\ (spectrogram)} \\
\hline
Fake                            &7.23 $\pm$\ 0.003          & -                 & -  \\
\hline
Gradient-based                  &3.96 $\pm$\ 0.203          & 0.00489     & 0.9618  \\
\hline
Inverse Mapping Model (ours)    &7.50 $\pm$\ 0.288          & 0.00196     & 0.9731  \\
\hline
Hybrid Method (ours)            &7.23 $\pm$\ 0.006          & 0.00003     & 0.9979  \\
\hline
\end{tabular}
\end{center}
\end{table*}

\begin{figure}[t]
\centerline{\includegraphics[width=1\linewidth]{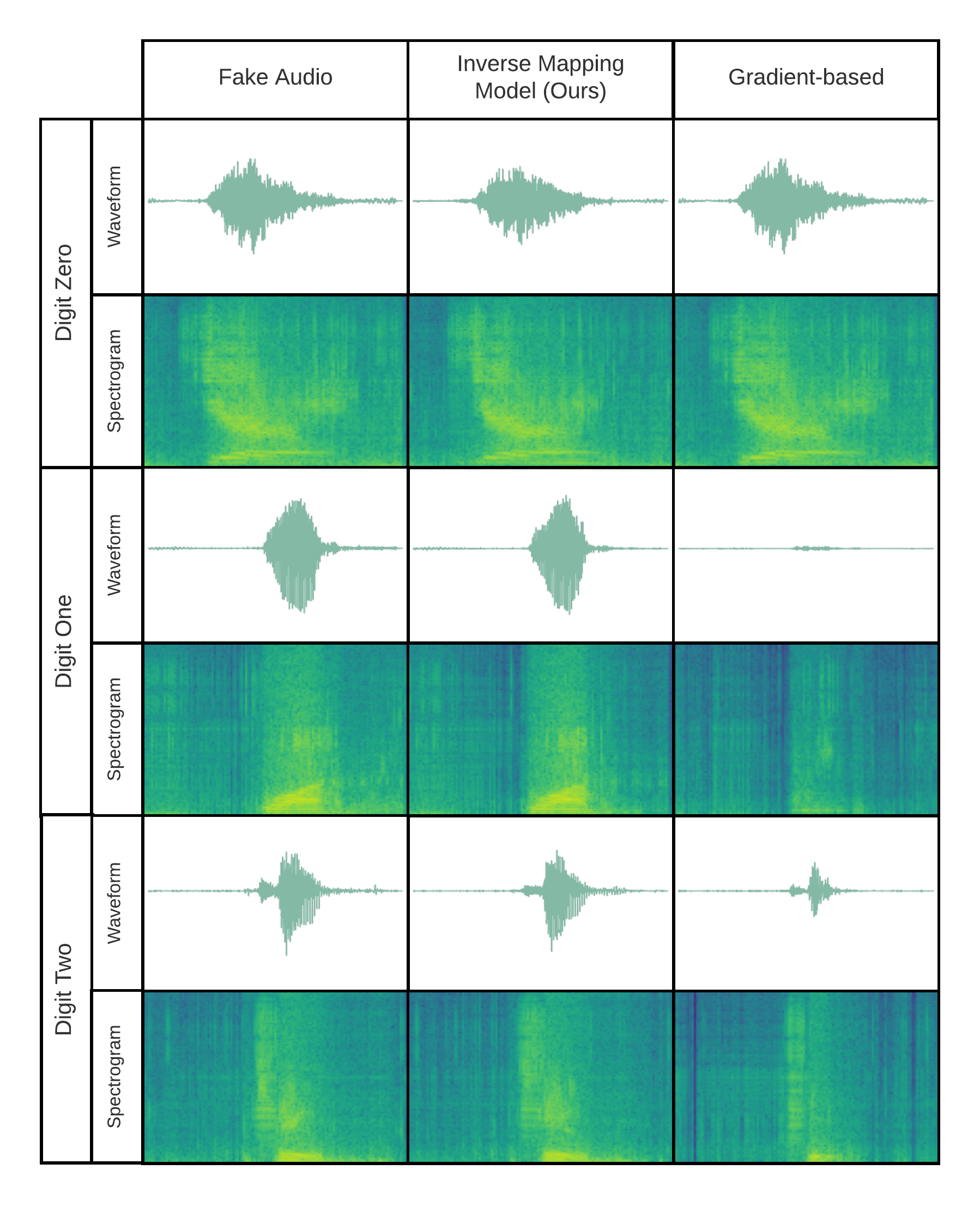}}
\caption{Comparing the Performance of Deep Network Inverse Mapping Model with Gradient Descent Approach in Latent Vector Recovery of Fake Audio. Three Example Digits are Shown Here. \label{fig2}}
\end{figure}
\begin{figure}[t]
\centerline{\includegraphics[width=1\linewidth]{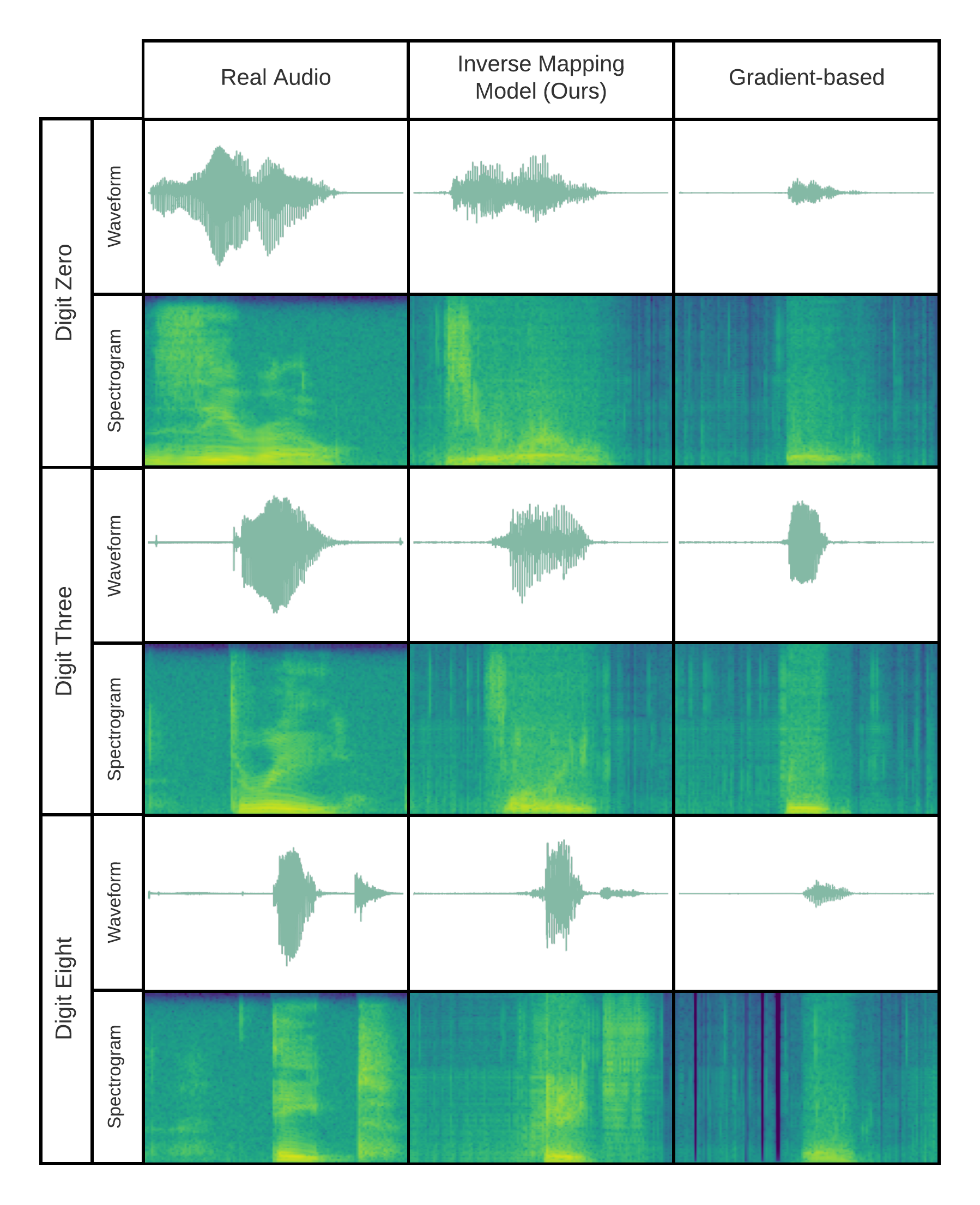}}
\caption{Comparing the Performance of Deep Network Inverse Mapping Model with Gradient Descent Approach in Latent Vector Recovery of Real Audio. \label{fig3}}
\end{figure}

\section{EXPERIMENTS AND RESULTS}
In this section, we compare the inverse mapping ResNet-18 model with sole gradient descent and hybrid methods based on MSE distance, SSIM measure and digit classification accuracy.
\subsubsection{Dataset}
Our generated audio dataset was created at run time by generating a random latent vector with a uniform distribution between -1 and 1. The latent vector was used as input to WaveGAN to generate raw-waveform audio. For real audio, we used the SC09 dataset which contains spoken digits from zero to nine. The training set includes 18620 samples with roughly equal occurrences of each digit by many different speakers.

\subsection{Implementation Details}
The inverse mapping model was trained for 250 epochs with an Adam optimizer using the learning rate of 0.001 and a batch size of 64. The only modifications to the ResNet-18 architecture was the removal of the final activation function and the number of classes being set to the size of WaveGAN's latent space. During each epoch, the model was first trained on one batch of real audio with perceptual loss and then one batch of WaveGAN synthesized audio with MSE between latent vectors and perceptual loss between the original audio and reconstructions. 

\subsection{Gradient Descent}
In order to recover the latent representation of a given audio using this method, we first sample a random vector, $z^*$, from the uniform distribution. Then we feed this latent vector to WaveGAN to get the corresponding audio. Next, we convert the audio to a spectrogram and compute the Mean Absolute Error (MAE) loss between the generated spectrogram and the target spectrogram. Afterward, $z^*$ is updated by the SciPy optimize library using the L-BFGS algorithm \citep{liu1989limited}. After each iteration, the updated $z^*$ is closer to the ground truth latent vector. Equation \ref{eq1} depicts the optimization process. $\mathcal{W}$ stands for pre-trained WaveGAN and $\mathcal{S}$ converts audio to spectrogram. The main disadvantage of the optimization methods is that they are very slow. Numerous iterations are required to find the optimum latent vector for a target sample; further, finding the perfect hyperparameters (learning rate and number of iterations) is extremely challenging. To recover latent vectors of real audio, we limited $z^*$ to 50000 gradient steps.
\begin{eqnarray}
min_{z^*} = \frac{1}{n} \sum|| \mathcal{S}(\mathcal{W}(z)) - \mathcal{S}(\mathcal{W}(z^*)) ||_1 \label{eq1}
\end{eqnarray}

\begin{table*}[h]
\caption{Real Audio Reconstructions} \label{real_audio_table}
\begin{center}
\begin{tabular}{|c|c|c|c|c|}
\hline
\textbf{}  & \makecell{Inception Score \\ (mean $\pm$\ std)} &  \makecell{MSE \\ (raw audio)} & \makecell{SSIM \\ (spectrogram)} & Accuracy \\
\hline
Real                            &9.20 $\pm$\ 0.019      & -             & -        & 95.41\% \\
\hline
Gradient-based                  &2.20 $\pm$\ 0.063      & 0.00959       & 0.953    & 17.17\% \\
\hline
Inverse Mapping Model (ours)    &7.93 $\pm$\ 0.077      & 0.01176       & 0.946    & 71.06\% \\
\hline
Hybrid Method (ours)            &3.89 $\pm$\ 0.027      & 0.00887       & 0.954    & 36.93\% \\
\hline
\end{tabular}
\end{center}
\end{table*}

\subsection{Experiment 1: Audio Reconstruction Using Inverse Mapping Model}
An initial objective of this project was to map generated audio back to latent space. The purpose of Experiment 1 was to evaluate the performance of inverse mapping model trained using pixel and perceptual loss in latent vector recovery of synthesized audio. The results in Table \ref{fake_audio_table} show we were able to recover the latent vector of synthesized audio with high accuracy. Since the gradient-based method is optimized on the MAE of the spectrograms, the metrics may appear to show a better reconstruction through gradient-based optimization. In actuality, the reconstruction through our inverse mapping model produces a more coherent spoken digit with better content and style reconstruction. This can be shown through the inception score of our inverse mapping model approach and is shown in Experiment 3 with real audio maintaining its digit classification. The qualitative results of our approach on fake audio are shown in Figure \ref{fig2} where our method is compared to gradient optimization for recovering the raw waveform and spectrogram of the audio. The same comparison on real audio is presented in Figure \ref{fig3}.

\subsection{Experiment 2: Comparing Different Approaches}
As mentioned earlier, optimization methods are capable of updating a random latent vector to recover the desired latent vector that reconstructs the target audio. We implemented an optimization-based method that updates a random vector for a maximum of 50000 steps using gradient descent. Finding the perfect learning rate for optimization-based approach is extremely challenging. In addition, we developed a hybrid method that first predicts a latent vector using our inverse mapping model and then updates it further with a maximum of 200 steps of gradient descent. The comparison between these three methods is summarized in Table \ref{fake_audio_table} for fake audio. Although the hybrid method produces a lower loss for the reconstruction of raw-waveform and spectrograms, the reproduced audio is of worse auditory quality. This indicates the inverse mapping model captures the important style and content of the audio while discarding noise. 

Overall, these results indicate that our inverse mapping model performs better than optimization-based methods.

\subsection{Experiment 3: Real Audio Reconstruction}
Very little was found in the literature on how to reconstruct real audio using GAN inversion techniques. A major problem with mapping real audio is that there are no ground truth latent vectors that accurately generate them. Another issue is the limitation of the audio GANs themselves, they are not capable of generating high quality naturalist audio that sound identical to the real peers. Nevertheless, our proposed approach is universal and can later be applied to invert more advanced GANs. The results of comparing our inverse mapping model with the alternatives on projecting real audio to latent representations is summarized in Table \ref{real_audio_table}. What stands out in the table is that while gradient-based methods reconstruct the waveform and spectrograms well, in terms of classification accuracy, our inverse mapping model is significantly better and is closest to the real audio accuracy. 

Together these results provide important insights into the latent vector recovery in the audio domain.

\section{CONCLUSION}
The present research aimed to examine different approaches of inverse mapping audio GANs. We proposed a deep residual neural network based approach to project both synthesized and real audio into latent space. We trained a ResNet-18 architecture based on a combination of reconstruction and perceptual losses. This study has identified that deep network solutions are not only faster than optimization-based alternatives, but also more accurate in terms of both auditory quality of the reconstructed sounds and digit classification accuracy.

Taken together, these findings suggest a role for deep neural networks in promoting inverse mapping of audio GANs. Being limited to current audio GANs, this study lacks high quality reconstructed audio for real sounds.

\subsubsection*{Acknowledgements}
The authors would like to thank Western BrainsCAN for the generous support of this research. The study was conducted on Compute Canada resources. A.K is supported by the Vector Scholarship in Artificial Intelligence, provided through the Vector Institute. 

\bibliographystyle{plainnat}
\renewcommand{\bibname}{References}
\renewcommand{\bibsection}{\subsubsection*{\bibname}}
\bibliography{arxiv}

\end{document}